# Integration of hBN quantum emitters in monolithically fabricated waveguides

Chi Li[1], Johannes E. Fröch[1], Milad Nonahal[1], Thinh N. Tran[1], Milos Toth[1,2], Sejeong Kim[3,*], Igor Aharonovich[1,2,*]

1. School of Mathematical and Physical Sciences, Faculty of Science, University of Technology Sydney, Ultimo, New South Wales 2007, Australia

2. ARC Centre of Excellence for Transformative Meta-Optical Systems (TMOS), University of Technology Sydney, Ultimo, New South Wales 2007, Australia

3. Department of Electrical and Electronic Engineering, University of Melbourne, Melbourne, Victoria 3010, Australia

**ABSTRACT**

Hexagonal boron nitride (hBN) is gaining interest for potential applications in integrated quantum nanophotonics. Yet, to establish hBN as an integrated photonic platform several cornerstones must be established, including the integration and coupling of quantum emitters to photonic waveguides. Supported by simulations, we study the approach of monolithic integration, which is expected to have coupling efficiencies that are ~ 4 times higher than those of a conventional hybrid stacking strategy. We then demonstrate the fabrication of such devices from hBN and showcase the successful integration of hBN single photon emitters with a monolithic waveguide. We demonstrate coupling of single photons from the quantum emitters to the waveguide modes and on-chip detection. Our results build a general framework for monolithically integrated hBN single photon emitter and will facilitate future works towards on-chip integrated quantum photonics with hBN.

**INTRODUCTION**

Over the last decade, hexagonal boron nitride (hBN) has emerged as a versatile material platform for a plethora of applications in nanosciences[1-3]. hBN was initially utilised as a "passive" material, where it became the standard choice for encapsulation of other 2D materials. Due to its atomically smooth surface and large bandgap, hBN facilitates higher carrier mobilities and minimal disturbance in the encapsulated materials[4]. Recently, hBN became an attractive material for nanophotonics, particularly in the fields of phonon polaritons and quantum technologies[5-8]. hBN is a host for a myriad of bright photoluminescent defects, which have potential for on-demand generation of single photons[9-14] and some of which exhibit optically detected magnetic resonance[15-20].

Integration of single photon emitters (SPEs) with photonic circuits is crucial as it enables scalable components packaged on a single chip. In this regard, it is important not only to generate SPEs on a chip, but also to demonstrate routing of single photons, which is commonly achieved by photonic waveguides[21-24]. For layered Van der Waals (vdW) materials, the most common approach has been the hybrid one where the vdW host of the quantum emitters is positioned on top of a foreign cavity fabricated from a different material[25-31]. As such, hBN emitters have been coupled to silicon nitride microdisks[32], waveguides and cavities[33-36] However, the monolithic approach, whereby the photonic components and the emitters are fabricated from the same material, can provide substantial benefits such as higher coupling efficiencies and less stringent fabrication steps[37].

Several optical components have been engineered from hBN so far, including metalenses, basic optical cavities and phononic resonators[38-42]. In this work, we build on these earlier works and demonstrate on-chip integration of hBN SPEs with a monolithic hBN waveguide. First, we elucidate the advantages of a monolithic approach over hybrid integration by comparing coupling efficiencies of selected configurations using photonic simulations. Then we fabricate on-chip hBN waveguides and SPE creation, enabling the coupling of hBN SPE to a waveguide mode and routing towards an on-chip grating couplers. Our results build the framework for further studies of monolithically integrated quantum light sources from vdW materials, and specifically hBN.

### RESULTS AND DISCUSSION

A general schematic of the monolithic waveguide is shown in **Fig. 1a**, where an optically active defect is embedded in the centre of the material, which also provides the waveguiding structure. The specific geometry that we study in this work is an hBN slot waveguide on an $SiO_2$ (285 nm)/Si substrate, with the geometry defined by a beam width of 1 μm, a height of 200 nm, and slot width of 80 nm. Typically, a slot waveguide is used to tightly confine the guided mode. In addition, this geometry is advantageous for our experimental scheme as it increases the probability to create SPEs near the centre of the waveguide. For a benchmark comparison, we first studied this particular structure considering different coupling schemes using the 3D finite-difference time-domain (FDTD) method. For all cases, we assumed a linearly polarized (x-pol) dipole emitter with a centre wavelength of 580 nm, located in or on top of the waveguide structure. As an example, the field distribution of a monolithically integrated SPE is shown in **Fig. 1b**. A fraction of the emission couples to the waveguide modes and propagates along the structure to grating couplers that scatter the light into the far field.

In addition to the monolithic configuration, we also considered two others – hybrid and surface – as is shown in **Fig. 1c**. The hybrid case consists of a relatively thick object with an embedded dipole emitter atop the waveguide. The surface case is particularly relevant to the integration of monolayers, whereby the monolayer is transferred on top of the waveguide, with the dipole emitter placed at the surface of the waveguide. To evaluate the effectiveness of the coupling to the waveguide mode, we extracted the emitter-waveguide coupling efficiency ($β$), defined as the ratio between the optical power guided through the waveguide and the optical power of the dipole emitter. Furthermore, the grating efficiency ($η$) indicates the amount of light that is scattered from the waveguide mode at the grating coupler into free space. From the FDTD simulations, we extracted $β$ to be 0.04 (0.11), in the hybrid (surface) configuration. The coupling efficiency increases dramatically to 0.4 if the dipole emitter is located in the centre of the waveguide, i.e. the monolithic configuration.

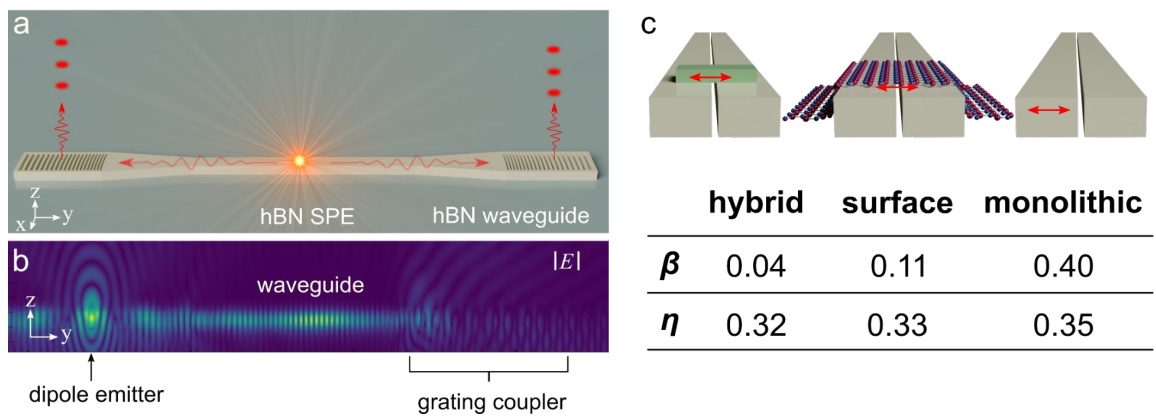

**Figure 1. FDTD simulations of an hBN emitter coupled to a waveguide.** *(a) Schematic illustration of a single photon emitter in a monolithic hBN waveguide. (b) 3D FDTD simulation of a monolithic hBN waveguide. (c) Schematics and table of the comparison of three situations, whereby the dipole is coupled from outside of the waveguide (hybrid), on the surface and embedded within the waveguide*

*(monolithic). The red arrows depict the corresponding dipole location and orientation of each case. β and η stand for emitter-waveguide coupling efficiency and grating coupler efficiency, respectively.*

In the hybrid case, the coupling efficiency can vary depending on the shape and the refractive index of the hybrid material. In this study, a rectangular block with 200 nm (width) × 200 nm (length) × 100 nm (height) with the same refractive index as the waveguide is simulated, where *β* reaches 0.04, which is on par with related works on hybrid coupled schemes[43]. The grating coupler used in this simulation extracts more than 30 percent of light into free space, with specific values for *η* of 0.32, 0.33, and 0.35 for the hybrid, surface, and monolithic cases, respectively.

To realise the SPE-integrated monolithic hBN waveguide concept, we fabricated structures based on the approach developed in our previous studies[38, 39], schematically depicted in **Figs. 2a-2d**. Further details of the fabrication are mentioned in the method section. An optical image of a set of photonic devices is shown in **Fig. 2e,** where the green area is a hBN flake. The same area was imaged by SEM, as is shown in **Fig. 2f**, and the individual waveguide is shown in **Fig. 2g**. The zoomed-in image shows a slot waveguide (**Fig. 2h**) and a grating coupler (**Fig. 2i**). For the above fabricated devices, the in and out coupling through the grating and guiding through the waveguide can be directly visualized by placing the laser spot in a confocal microscope setup at one grating coupler and monitoring a CCD camera image. Such an image is shown in **Fig. 2j** with the outline of the waveguide drawn for clarity.

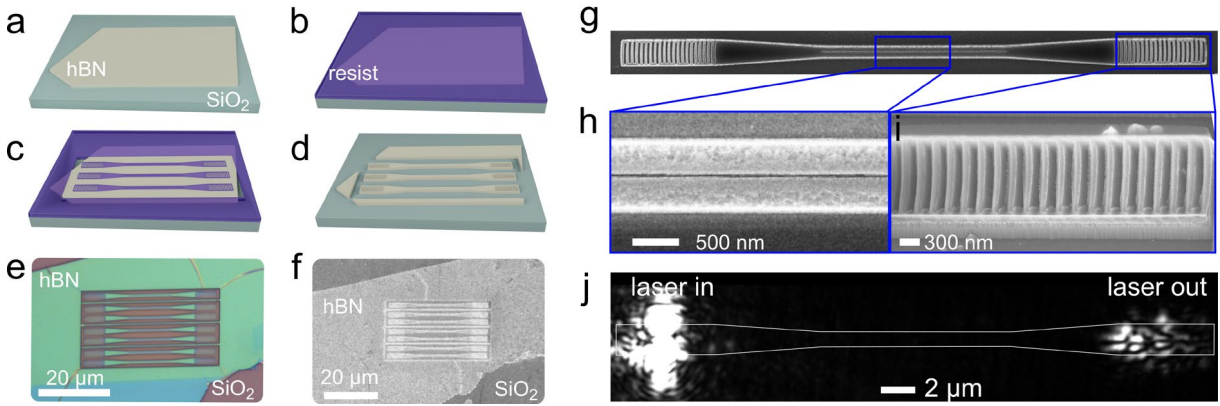

**Figure 2**. *(a)-(d) Fabrication of monolithic hBN waveguides: (a) is exfoliation of hBN flakes on clean $SiO_2$ substrate, (b) spin coating with e-beam resist, (c) EBL and development processes, (d) RIE etch and resist removal. (e) and (f) are optical and SEM images of one fabricated sample. (g)-(i) SEM images of a waveguide and a zoomed-in view of a slot waveguide and a grating coupler component, respectively. (j) widefield image of a laser being coupled and guided through the waveguide.*

In the following, we identified SPEs in the fabricated devices by standard confocal PL mapping as shown in **Fig. 3a** (the emitters were generated using a plasma processing and an annealing treatment, as is detailed in methods). Here, an emitter (marked by a yellow star) was found at the centre of the monolithic hBN waveguide (outlined by white lines). The spectrum of the SPE (**Fig. 3b**) displays a prominent zero phonon line (ZPL) around 590 nm with a phonon sideband (PSB) at 640 nm, exhibiting the characteristic detuning of hBN emitters of 164 meV. In addition, the narrow peak around 573 nm corresponds to the hBN $E_{2g}$ Raman peak under 532-nm excitation. The second-order correlation function (inset), collected with a spectral filter (bandwidth is indicated with light blue shading in the spectra) confirmed the quantum nature of the emitter with a dip at $g^{(2)}(0)$ of ~ 0.4. We note that the measurement was not corrected for background.

To verify that the emission is coupled to the hBN waveguide, we fixed the laser excitation spot at the position of the SPE, while scanning the collection spot, using a decoupled excitation and collection paths in the PL setup. The respective PL map is shown in **Fig. 3c** with the brightest spot at the centre (SPE) corresponding to the same excitation/collection spot. At the same time, higher PL intensities were measured from the couplers, denoted as spots C1 and C2, which is direct evidence for coupling to the waveguide mode and directional scattering at the grating couplers. The spectra collected from the positions C1 and C2 are shown in **Figs. 3d and 3e**, respectively. The same ZPL is indeed shown at 590 nm, further confirming that the emission is coupled to the waveguide mode. The periodic intensity variation with a periodicity of ~ 5 nm superimposed on the spectra collected from the grating coupler likely corresponds to a Fabry-Perot mode of the waveguide, determined by the waveguide length of 25 µm. Whilst the data is from the same single emitter, the $g^{(2)}$ curves shown in insets differ – specifically, the value of $g^{(2)}(0)$ is slightly higher when collected from the grating coupler. This is because of the increased collection time for $g^{(2)}$ due to reduced ZPL intensity and an overall lower signal to background ratio at the grating coupler.

To determine the coupling efficiency, we measured the emitter polarization relative to the waveguide, shown in a polar plot in **Fig. 3f**. The radial axis is normalized APD counts of emission intensity from 0 to 1. First, the waveguide longitudinal orientation is set to 0º as shown by the black line with arrows. The red and blue circles are the emission and absorption intensity, respectively. The corresponding red and blue dashed curves are fits. Accordingly, the red line with an arrow is the emitter polarization direction, which yields an orientation of 70 º relative to the longitudinal direction of the waveguide. Therefore, based on our prior simulation a maximum value of 0.37 is achievable for the emitter-waveguide coupling efficiency ($β$). To experimentally estimate the coupling efficiency, we compared the relative intensity measured directly from the SPE location and from the grating coupler C2. When both spectra were measured with the same excitation power and integration time, the relative intensity from the position of the SPE is 1150 cps while the intensity at C2 is 81 cps. Considering the out-coupling efficiency of the grating coupler of 0.35, we obtain the relative intensity into one arm of the waveguide of ~ 231 cps. Furthermore, from simulation, we obtained that only 0.08 of the intensity of the SPE is otherwise collected, which yields a value of 14400 cps as the absolute emitter intensity. This allows us to deduce the emitter-waveguide coupling efficiency which is 231×2/14400=0.032. We note that the coupling efficiency is lower than the theoretically predicted value, which may be attributed to the lower grating efficiency and higher waveguide loss due to imperfections in nanofabrication.

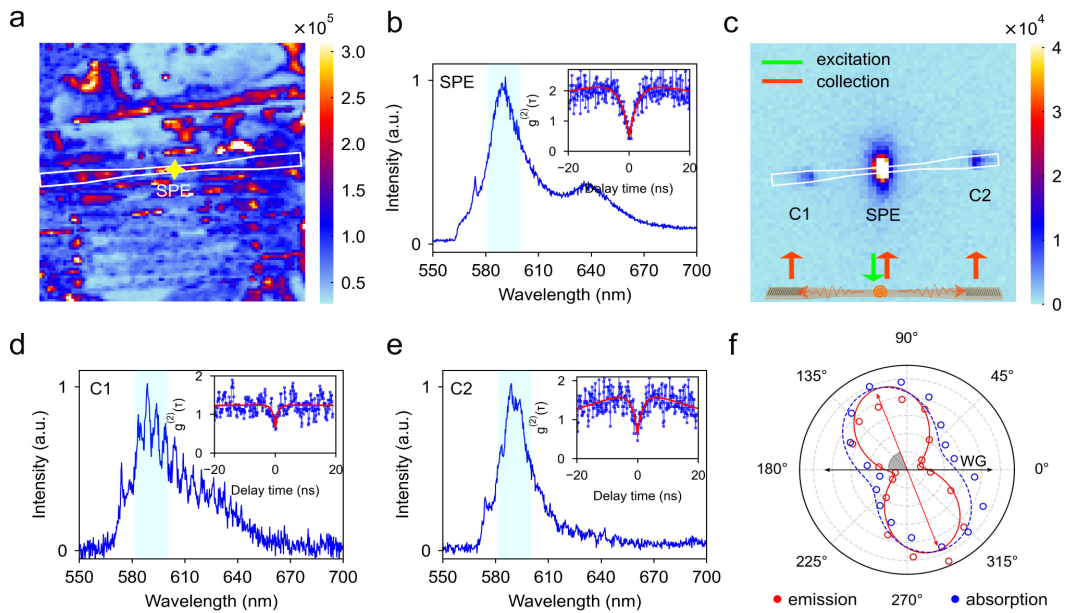

**Figure 3**. *An emitter coupled to a waveguide.* (a) Stage scanning confocal image of hBN waveguides. Note that excitation and collection were from the same spot in this case. The emitter and waveguide are

*marked in the map by a yellow star and white line respectively. (b) SPE spectrum. The inset is the antibunching curve. The excitation power was 500 μW. (c) Collection images when the laser spot is fixed on the emitter. C1 and C2 denote the couplers at the two ends. The inset is the corresponding schematic for excitation and collection. (d), (e) are corresponding spectra and antibunching curves from C1 and C2 sites, respectively. The excitation power was 500 μW. (f) Emitter emission intensity vs polarizer angles (red curve) and doubled angles of the half wave plate on the excitation path (blue curve). The black arrow shows the waveguide longitudinal orientation, while the red arrow shows the emitter dipole orientation. The intersection angle is about 70º.*

To prove the capabilities of our design and fabrication process we demonstrated further key elements that are needed for an integrated photonics platform. Specifically, examples of various configurations fabricated with the recipe detailed in the methods section are shown in **Fig. 4**.

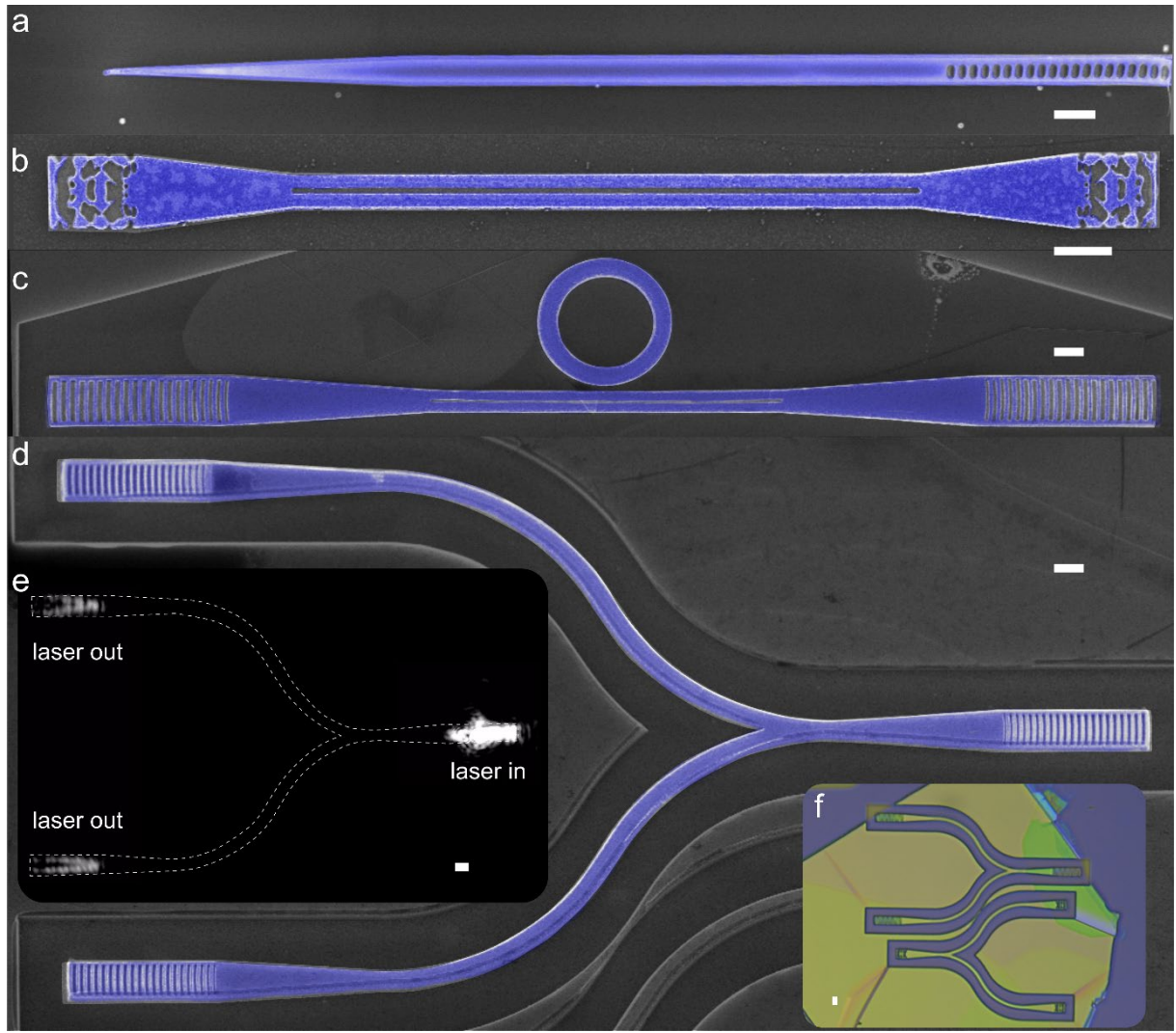

**Figure 4**. *False coloured SEM images of other hBN monolithic waveguides. (a) a tapered waveguide (b) a waveguide with inverse couplers (c) a grating waveguide with a microring resonator (d) a beam splitter with grating couplers. (e), (f) optical images of bright field and a laser coupling demonstration of the splitter, respectively. The scale bars in all images correspond to 1 um.*

Here, **Fig. 4a** shows a tapered waveguide - a configuration which is particularly advantageous for hybrid integration[23, 44], as well as for side collection schemes[37, 45]. For the former use-case the taper enables the adiabatic coupling in a hybrid scheme, where the waveguide element is integrated on top of another waveguide. For the latter use-case the tapering facilitates a more directional coupling. **Fig. 4b** shows an example of a waveguide with two inverse designed grating couplers[46]. Here the grating couplers were designed using the Lumerical API automation. A further example of a functionalized waveguide is shown in **Fig. 4c**, where a waveguide is placed next to a microring. Such a scheme can be utilized for to either enhance the emission of an integrated colour centre[47], or as a narrow-band filter[48]. In the former case, an SPE that is integrated within the microring experiences enhancement due to the Purcell effect as it couples to the resonator mode, determined by the quality factor and the mode volume. To achieve the functionality of an on-chip narrowband filter, a second waveguide, placed on the opposite side of the waveguide would permit only emission of the SPE that couples to the microring mode to transmit to the other waveguide. Finally, **Fig. 4d** shows an hBN waveguide with 50:50 beam splitter, which can be utilised for on-chip Hanbury Brown-Twiss (HBT) measurements. **Fig 4e** and **Fig 4f**, illustrate optical images of a splitter sample, where a green laser was coupled to demonstrate the functionality. We emphasise that our fabrication conditions yield clear outlines, straight sidewalls and well-resolved features expected in the structures. In particular, small feature sizes in the inverse design coupler are achieved.

**CONCLUSIONS**

In conclusion, we presented the fabrication of various hBN monolithic waveguides – including a coupled ring resonator, splitter and a straight waveguide. We achieved and demonstrated monolithic integration of SPEs in hBN waveguides with a coupling efficiency of 0.032. In simulations, we demonstrated the advantage that a higher coupling efficiency can be achieved by employing a monolithic approach as compared to hybrid integration. Our work contributes to future studies of monolithically-integrated hBN SPEs for waveguide integration and on-chip manipulation.

**ACKNOWLEDGEMENTS**

The authors would like to thank Simon White and Minh Nguyen for their kind help on the optical setup maintenance. We acknowledge the Australian Research Council (CE200100010) and the Asian Office of Aerospace Research and Development (FA2386-20-1-4014).